# Instability of Amorphous Oxide Semiconductors via Carrier-Mediated Structural Transition between Disorder and Peroxide State

Ho-Hyun Nahm[1], Yong-Sung Kim[1,*], and Dae Hwan Kim[2]

[1]Korea Research Institute of Standards and Science, Yuseong, Daejeon 305-340, Korea

[2]School of Electrical Engineering, Kookmin University, Seoul 136-702, Korea

[*]e-mail: yongsung.kim@kriss.re.kr

**Abstract**

The excited holes occupying the valence band tail states in amorphous oxide semiconductors are found to induce formation of meta-stable $O_2^{2-}$ peroxide defects. The valence band tail states are at least partly characterized by the O-O $pp\sigma^*$ molecular orbital, and the localized-hole-mediated lattice instability results in the formation of the peroxide defects. Along with the O-O bond formation, the $pp\sigma^*$ state is heightened up into the conduction bands, and two electrons are accordingly doped in the electronic ground state. The energy barrier from the $O_2^{2-}$ peroxide state to the normal disorder state is found to be 0.97 eV in hybrid density functional theory. The hole-mediated formation of the meta-stable peroxide defects and their meta-stability is suggested as an origin of the negative bias and/or illumination stress instability in amorphous oxide semiconductors.

## I. Introduction

The limit of carrier mobility of amorphous Si makes it challenging to apply the amorphous Si-based thin film transistors (TFTs) to next-generation flat panel displays, and the amorphous oxide semiconductors (AOS), which have about ten times larger electron mobility, have attracted great attention as a material to replace the amorphous Si.[1,2] The AOS material is not only suitable for low temperature processing but also visible-light transparent and flexible.[3,4] Thus, extensive studies on the AOS materials and TFTs have been made to understand the material and device properties including a variety of puzzling phenomena and realize the display applications.[1-4] However, there is a critical unsolved problem of the negative bias and/or illumination stress instability. The material and device properties, such as conductivity and threshold voltage ($V_{th}$), are significantly changed and recovered at a very slow rate after the stress.[5-16] This instability problem is now a critical obstacle to fabricating reliable AOS-based devices.

There has been considerable effort to resolve the negative bias and/or illumination stress instability of AOS. (i) Holes are generated by light illumination and trapped at the interface between the AOS channel and the gate insulator and/or injected into the gate insulator.[8,14] (ii) Surface O atoms are desorbed (reduction) and accordingly electrons dope.[7] (iii) Photo-ionized O-vacancies ($V_O$) persist with slow electron recombination.[9-14,17,18] Although the gate insulator hole trap model (i) can be one of the instability mechanisms, the instability has been observed not only in AOS TFTs but also in AOS thin films without a gate insulator through persistent photo-conductivity (PPC).[14,19] Thus, there has been considered to be an intrinsic mechanism in AOS materials.[12,14] According to the surface O reduction model (ii), an overlying passivation layer on the AOS TFTs has been adopted,[7] but the instability has been still serious even avoiding the surface effect.[8-10,12] The $V_O$ (iii) can give photo-carriers and be a hole-trap,[9-14,17,18] but the meta-stability of the ionized $V_O$ remains unclear. In order to explain the long recovery time of $V_{th}$ in AOS TFTs[9,14-16] and the long PPC of AOS[14,19] (an order of day at room temperature), the energy barrier from the meta-stable state to the stable state should be very high. Experimentally reported activation energy is in the range of 0.9~1.0 eV.[19]

In this work, we show that the carrier-mediated structural transition between the normal disorder and the $O_2^{2-}$ peroxide state leads to the negative bias and/or illumination stress instability of AOS. With excited holes in the valence band tail (VBT) states, the empty $pp\sigma^*$-like VBT states induce a driving force to form an O-O bond through the $pp\sigma$-hybridization. The $pp\sigma^*$ state is heightened up into the conduction bands along with the O-O bond formation, and in the electronic ground state the two electrons occupy the conduction band minimum (CBM) state. The $O_2^{2-}$ peroxide state is found to be meta-stable, and the energy barrier from the peroxide state to the normal disorder state is calculated to be 0.97 eV.

## II. Calculational Method

Density-functional theory (DFT) calculations were performed as implemented in the Vienna ab initio simulation package (VASP) code.[20,21] The projector-augmented wave (PAW) pseudopotentials[22] and the plane-wave basis set with a kinetic energy cutoff of 400 eV were used. The local-density approximation (LDA)[23] with the on-site Coulomb energy (U) of 6 eV for the In, Ga, and Zn d states[24] (LDA+U) was mainly used for the exchange correlation energy. The Heyd-Scuseria-Ernzerhof (HSE) hybrid functional[25,26] was also used with the standard mixing parameter of 0.25 (25% from Hartree-Fock) and the screening parameter fixed to 0.2 Å$^{-1}$. The LDA+U band gap of the amorphous $InGaZnO_4$ is 0.97 eV, and the HSE band gap is 2.55 eV similarly to the other HSE calculations.[27] The HSE value is more close to the experimental (optical) band gap of 3.17 eV.[28] There were only quantitative differences between the LDA+U and HSE calculations.

We mainly investigated amorphous $InGaZnO_4$, which is a representative AOS, and $In_2Ga_2ZnO_7$, $In_2Sn_2Zn_2O_9$, and $In_2Sn_6Zn_{12}O_{27}$ AOS's were also studied.[29] The 168-atomic cubic supercell was used for the $InGaZnO_4$ [Fig. 1(a)], and the 144-, 180-, 141-atomic cubic supercells were used for the $In_2Ga_2ZnO_7$, $In_2Sn_2Zn_2O_9$, and $In_2Sn_6Zn_{12}O_{27}$, respectively. The average volumes per unit formula of the AOS's were derived from the experimental mass densities of $In_2O_3$ (7180 kg/m$^3$), $Ga_2O_3$ (6000 kg/m$^3$), $SnO_2$ (6900 kg/m$^3$), and ZnO (5600 kg/m$^3$). The unit lengths of the cubic supercells are 12.542 ($InGaZnO_4$), 11.895 ($In_2Ga_2ZnO_7$), 13.047 ($In_2Sn_2Zn_2O_9$), and 11.970 Å ($In_2Sn_6Zn_{12}O_{27}$). For $InGaZnO_4$, the amorphous phase is 9.6 % larger in average volume per unit formula than the crystalline phase. The orthorhombic and rhombohedral supercells were also tested for the amorphous $InGaZnO_4$. The 168- and 112-atomic supercells were used, respectively [Figs. 1(b) and 1(c)]. No meaningful differences were found between the cubic, orthorhombic and rhombohedral supercells.

A single k-point at (0.25, 0.25, 0.25) in the cubic Brillouin zone (BZ) was used in most calculations for the BZ integration. The total energy difference between the single k-point and the 2×2×2 Monkhorst-Pack mesh is only a few meV. In β evaluation (will be defined below), we used the Γ point to describe the electrons occupying the CBM state. In α (will be defined below) and relative stabilities, the energy differences between the (0.25, 0.25, 0.25) and Γ point calculations are only less than 10 meV.

The amorphous structures were generated by the melt-and-quenching molecular dynamics simulations within the Nosé canonical ensemble. A random AOS structure was melted at 3000 K for 6 ps and then quenched down to 300 K with the quenching rate of 900 K/ps. The amorphous structures were then optimized by the static DFT calculations until the Hellmann-Feynman forces were less than 0.02 eV/Å. For amorphous $InGaZnO_4$ in cubic supercell, the pair-correlation functions of the In-O, Ga-O, Zn-O, and O-O atoms are shown in Fig. 2. The average coordination numbers of the In, Ga, and Zn atoms are 5.42 (within r=2.5 Å), 4.58 (r=2.2 Å), and 4.29 (r=2.2 Å), respectively, with bonding to O atoms.

### III. Results
#### 1. Structural Transitions

Figure 3(a) shows a local atomic structure of the as-generated amorphous $InGaZnO_4$, and we denote this local structure as a disorder state (DS). In order to simulate the excited holes in the stress conditions, we put two holes into the supercell and investigate the structural changes in the amorphous InGaZnO$_4$. The disorder state (DS$^*$) in the (2+) hole-injected case is found to be meta-stable and prefers to be structurally transformed into the O$_2^{2-}$ peroxide state (PS$^*$), of which structure is shown in Fig. 3(c). We plot the calculated total energies (LDA+U) as a function of the O-O distance in Fig. 3(e) (red curve). The peroxide state (PS$^*$) is found to be 0.88 eV more stable than the disorder state (DS$^*$). The transition state (TS$^*$) is found as shown in Fig. 3(b), and the energy barrier (α) in the DS$^*$ to PS$^*$ transition is found to be 0.26 eV [Fig. 3(e)]. This reaction can be written as

$$O^{2-} + O^{2-} + 2h^+ \rightarrow O_2^{2-}, \quad (1)$$

and indicates that the peroxide is formed via mediated by holes.

After the stress, the AOS materials go back to the electronic ground state. In the neutral charge state, the O$_2^{2-}$ peroxide state (PS) is found to be meta-stable on the contrary to the (2+) hole-injected case. As shown in Fig. 3(e) (blue curve), the peroxide state (PS) is higher in energy by 1.25 eV than the disorder state (DS). That is the reason why the peroxide defects are not preferentially formed in the as-generated amorphous structures. (It has been reported that peroxides are formed in rapidly quenched amorphous structure.[27]) A significant energy barrier (β) of 0.69 (LDA+U) and 0.97 eV (HSE) is found in the PS to DS transition [Fig. 3(e)]. The atomic structure of the transition state (TS) is depicted in Fig. 3(d). The structural recovery process can be written as

$$O_2^{2-} + 2e^- \rightarrow O^{2-} + O^{2-}, \quad (2)$$

and the significant β energy barrier indicates that it is a very slow process and the long persistency of the meta-stable O$_2^{2-}$ peroxide defect.

## 2. Electronic Structures

In order to certify the hole-mediated formation of the peroxide defect ($DS^*$-to-$PS^*$ transition), we investigate the electronic structures. The local electronic densities of states (LDOS) near the two O atoms are shown in Fig. 4 for various O-O distances. In the (2+) hole-injected disorder state ($DS^*$), the Fermi level crosses the VBT states [Fig. 4(a)]. As the O-O distance is closer, a state emerges into the band gap with decoupled from the valence bands [Fig. 4(b)]. The state is found to be $pp\sigma^*$-like between the two O atoms [Fig. 5(a)]. At the transition state ($TS^*$), the $pp\sigma^*$-like state is almost empty [Fig. 4(b)]. As the O-O distance is furthermore closer, the $pp\sigma^*$ level goes up even higher, and in the $O_2^{2-}$ peroxide state ($PS^*$), it is found inside the conduction bands [Fig. 4(c)]. Since, in the (2+) hole-injected case, the $pp\sigma$ bonding state is largely stabilized as the $pp\sigma^*$ anti-bonding level rises up [Fig. 4(d)], the peroxide state ($PS^*$) becomes more stable than the disorder state ($DS^*$). The deoccupation of the anti-bonding $pp\sigma^*$-like VBT states (VBT holes) is the fundamental origin giving the driving force to form the peroxide defect.

The essential point here is that the VBT holes in AOS are characterized by the localized O-O $pp\sigma^*$ anti-bonding state [Fig. 5]. The reason why the VBT states have the $pp\sigma^*$ character is obvious. In AOS, the upper valence bands are characterized by the O-2p anti-bonding states. In inter-site p-p coupling, the $\sigma$-$\sigma^*$ level splitting is much larger than the $\pi$-$\pi^*$ splitting, and the $pp\sigma^*$ forms the highest energy level. Since the VBT states are the highest among the anti-bonding states, they become to have at least partly the O-O $pp\sigma^*$ anti-bonding character. We find similar $pp\sigma^*$ character of the VBT states in different amorphous structures of InGaZnO$_4$, and also in other AOS's of the In$_2$Ga$_2$ZnO$_7$, In$_2$Sn$_2$Zn$_2$O$_9$, and In$_2$Sn$_6$Zn$_{12}$O$_{27}$ (Fig. 5). It can be also found in previous other calculations.[30,31]

The meta-stability of the peroxide defect (PS-to-DS transition) is also clearly understood from the electronic structures. With the peroxide defect in the neutral charge state, the two electrons occupy the CBM state and the Fermi level is located near the CBM [Fig. 4(g)], where the pp$\sigma^*$ anti-bonding state is well empty. With breaking the O$_2^{2-}$ peroxide bond, the pp$\sigma^*$ level goes down with increasing the total energy (due to heightening up the occupied pp$\sigma$ state), and when the pp$\sigma^*$ state crosses the CBM, two electrons occupy again the pp$\sigma^*$ state [Fig. 4(f)]. The transition state (TS) is thus found when the pp$\sigma^*$ is at the CBM [Fig. 4(f)]. The O-O bond breaking is accelerated only after two electrons occupying the pp$\sigma^*$ state, and the total energy lowers as shown in Fig. 3(e), rendering the peroxide state (PS) back to the disorder state (DS). Thus, the β energy barrier originates from the level separation (δ) between the Fermi level and the unoccupied pp$\sigma^*$ level in the peroxide state [Fig. 4(g)]. The δ is calculated to be 2.9 (LDA+U) and 3.9 eV (HSE) above the CBM (Fig. 6). Due to the high pp$\sigma^*$ level in the peroxide state and the weak coupling with the conduction bands [there is no O-related state in the energy range of 1~3 eV in Fig. 4(g)], the β energy barrier is as high as 0.69 (LDA+U) and 0.97 eV (HSE), and it can be even higher if the Fermi level was inside the band gap (by some gap states).

## IV. Discussion

The carrier-mediated structural transitions have significant implications to the negative bias and/or illumination stress instability of AOS, and explain important experimental findings. We should emphasize that only if VBT holes are generated through any possible hole-generating stress, it is possible to form the peroxide defects. By negative bias and/or illumination stress, which generates electron-hole pairs, the excited VBT holes act to form the peroxide defects through

$$O^{2-} + O^{2-} + 2h^+ + 2e^- \rightarrow O_2^{2-} + 2e^-. \quad (3)$$

Due to the meta-stability of $O_2^{2-}$, the two electrons ($2e^-$) occupying the CBM state result in the increase of the Fermi level and the negative shift of $V_{th}$ in AOS TFTs for a long time. The calculated β energy barrier (0.97 eV in HSE) for the recovery (Eq. 2) is very close to the experimental activation energy of 0.9~1.0 eV.[19] In illumination stress, UV light should be very effective to generate VBT holes, and visible light can also induce VBT holes through the gap-state-assisted excitations [Fig. 7(a)], because there are lots of gap states in AOS.[31] Especially under combined negative bias and illumination stress (NBIS), the negative bias field accumulates the VBT holes near the interface between the AOS channel and the gate insulator [Fig. 7(b)]. Then, a large density of peroxide defects can be generated near the interface in proportional to the VBT hole density. The negative bias will also reduce the α barrier by electrostatic energy through α'=α-qV, where q is the effective charge and V is the applied voltage near the interface.[32,33] Thus, the NBIS can be very serious in AOS TFT instability. By only negative bias stress (NBS), it would be less effective to generate VBT holes, but if the applied voltage is large enough to form inversion AOS channel [Fig. 7(c)], the meta-stable peroxide defects can be formed near the interface.

In terminology of defects in crystalline semiconductors, the peroxide formation can be considered as a formation of a $V_O$ and O-interstitial ($O_i$) pair (O Frenkel-pair). The $O_i$ is in a split-interstitial form with bonding to a host O atom (O-$O_i$), and charge neutral [$(O^{2-}\text{-}O_i^0)=(O\text{-}O_i)^{2-}$]. The $V_O$ is a (2+) donor, and thus the two electrons can be considered to be doped by the created $V_O$. The $V_O$ is found to be a shallow donor in all our calculations, and the Fermi level is found at the CBM in the peroxide state, as shown in Fig. 4(g). Although we cannot exclude the formation of a deep $V_O$ donor, the O-O bond near the $V_O$ can hinder the inward relaxation of the cation atoms around the $V_O$ [Fig. 3(c)]. Experimentally, a reversible structural change that requires some relaxation time has been suggested and the generation and recombination of $V_O$ and $O_i$ have been conjectured, which are proven theoretically by our structural transition model.[34] The meta-stable peroxide defect is also similar to the double-broken-bond (DBB) acceptor-compensating (AX) center in acceptor-doped crystalline semiconductors.[35] Two cation-O bonds in the disorder state are broken and one O-O bond is created in the peroxide state, as shown in Figs. 1(a) and 1(c), respectively. The AX-like behavior of the peroxide defect can make p-type doping more difficult in AOS as an intrinsic compensating center.

From the point of view of material engineering, it is important fundamentally to reduce the density of localized VBT states in AOS to improve the stability, since the formation of the meta-stable peroxide defects is mediated by the VBT holes. Post-annealing process has been known to reduce the density of VBT states.[36] It is because of reducing the high energy local amorphous configurations. The minimization of the VBT states should be a direction in process and material designs to obtain reliable AOS's.

## V. Conclusion

The negative bias and/or illumination stress instability of AOS is suggested to originate from the carrier-mediated structural transition between the disorder and peroxide state. The peroxide state is meta-stable and the activation energy for the structural recovery is found to be 0.97 eV in hybrid density-functional theory. The main reason for the peroxide formation lies in the pp$\sigma^*$-like VBT states. Reduction of the VBT states is suggested as a fundamental direction to improve the stability of AOS-based devices.

**Acknowledgement**

Y.S.K. acknowledges the support by Nano R&D program through the National Research Foundation (NRF) of Korea funded by the Ministry of Education, Science, and Technology (MEST) (No. 2009-0082489). D.H.K. acknowledges the support by the NRF grant funded by the Korea government (MEST) (No. 2011-0000313).

**Figure Captions**

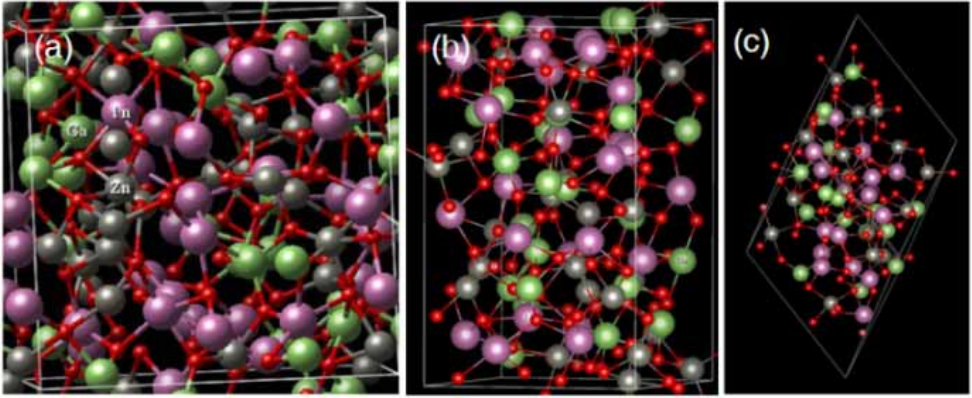

**Figure 1.** (Color online) Atomic structures of the as-generated amorphous InGaZnO$_4$ in (a) cubic, (b) orthorhombic, and (c) rhombohedral supercells. Large balls are cations and small (red) ones are O atoms.

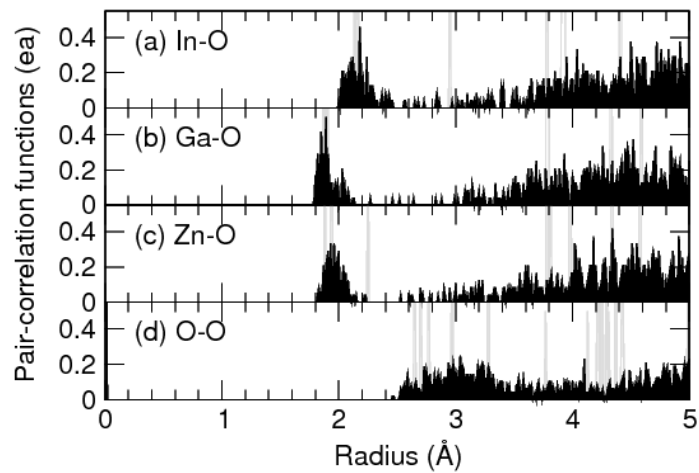

**Figure 2.** Pair-correlation functions of the (a) In-O, (b) Ga-O, (c) Zn-O, and (d) O-O atoms for the generated InGaZnO$_4$ amorphous structure (black filled areas). The gray lines indicate those for crystalline InGaZnO$_4$.

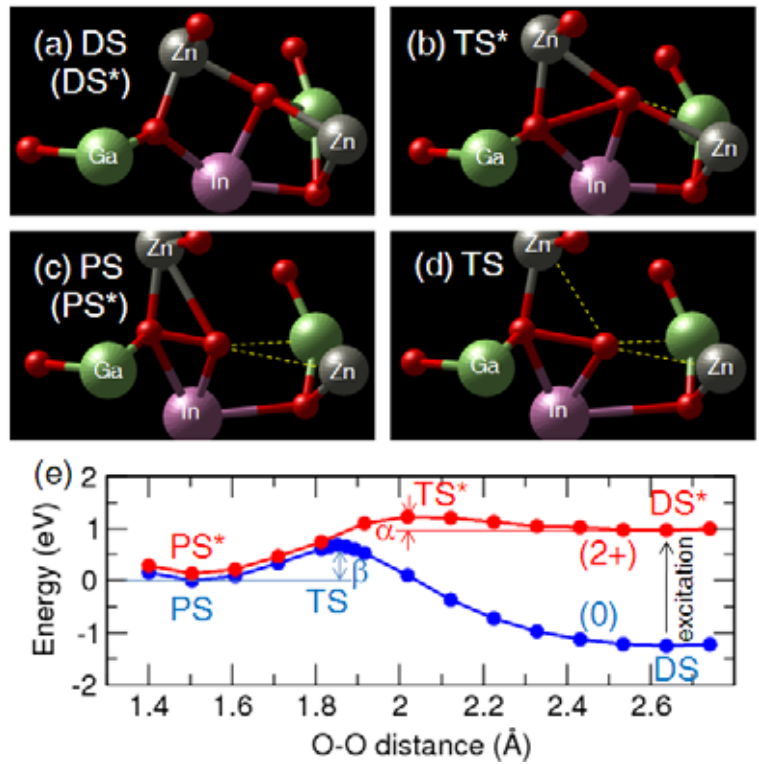

**Figure 3.** (Color online) Local atomic structures of amorphous InGaZnO$_4$ in (a) the disorder states (DS and DS$^*$), (b) (2+) charged transition state (TS$^*$), (c) peroxide states (PS and PS$^*$), and (d) neutral transition state (TS). Small (red) balls are O atoms, and large ones are cations. (e) Calculated total energies (LDA+U) as a function of the O-O distance in (0) and (2+) charge states, with respect to the PS energy.

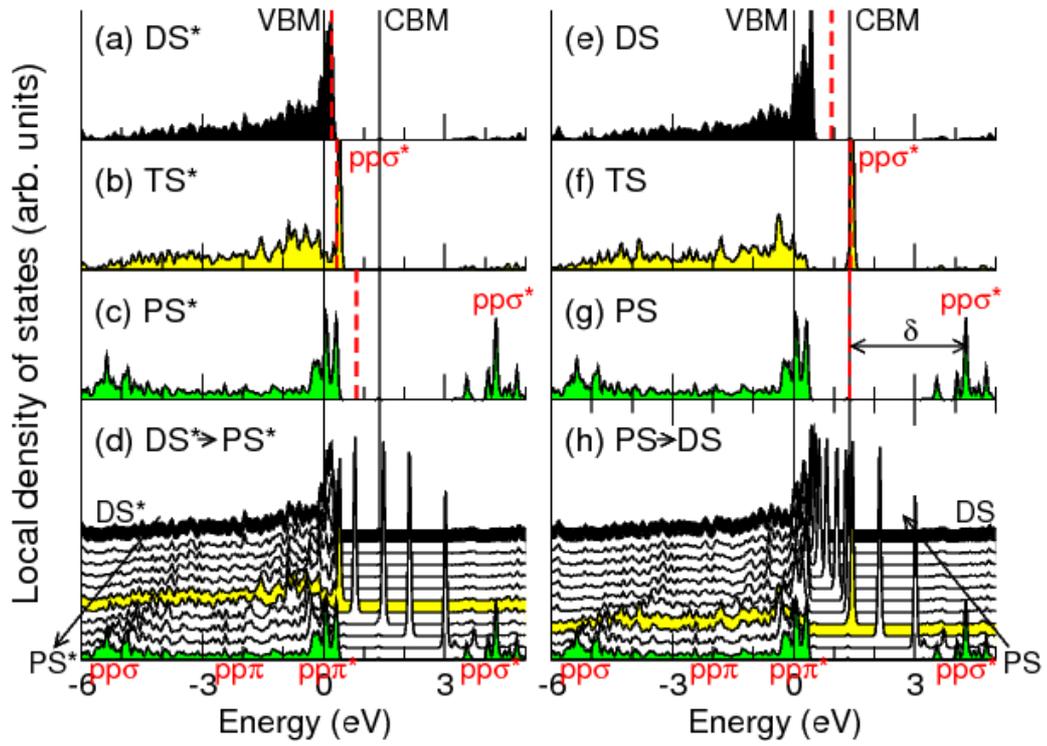

**Figure 4.** (Color online) Calculated LDOS (LDA+U) near the O pair in the (a) DS*, (b) TS*, and (c) PS* in the (2+) hole injected case. As the O-O distance is closer (DS*→PS*), the LDOS evolution is shown in (d). The LDOS in the (e) DS, (f) TS, and (g) PS in the neutral charge state. For the recovery (PS→DS), the LDOS evolution is shown in (h). The Fermi levels are indicated by vertical (red) dashed lines. The VBM is the valence band maximum of the crystalline $InGaZnO_4$.

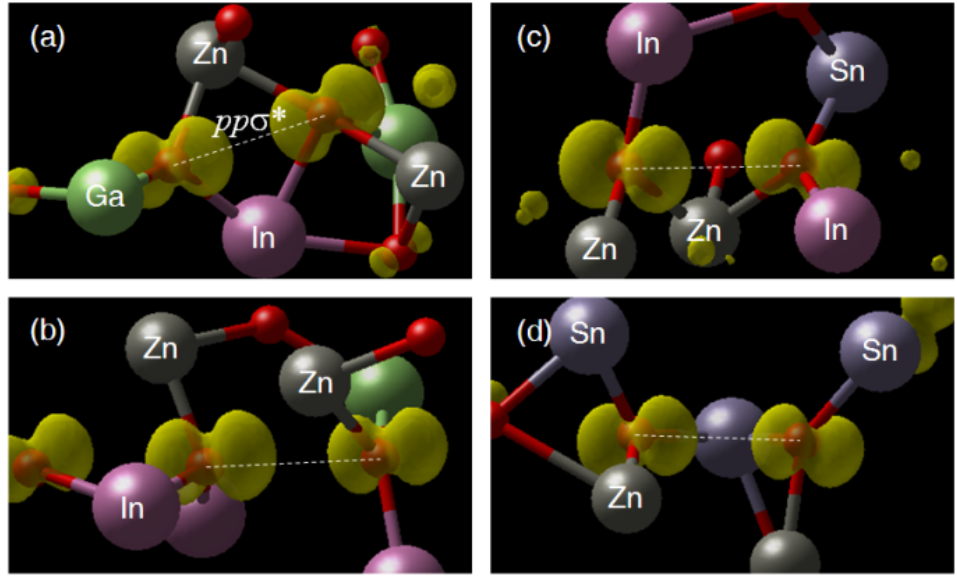

**Figure 5.** (Color online) Charge densities of the top-most VBT states of the amorphous (a) InGaZnO$_4$, (b) In$_2$Ga$_2$ZnO$_7$, (c) In$_2$Sn$_2$Zn$_2$O$_9$, and (d) In$_2$Sn$_6$Zn$_{12}$O$_{27}$. Large balls are cations and small (red) ones are O atoms. The dashed lines connect two O atoms having the pp$\sigma$* orbital in the VBT state.

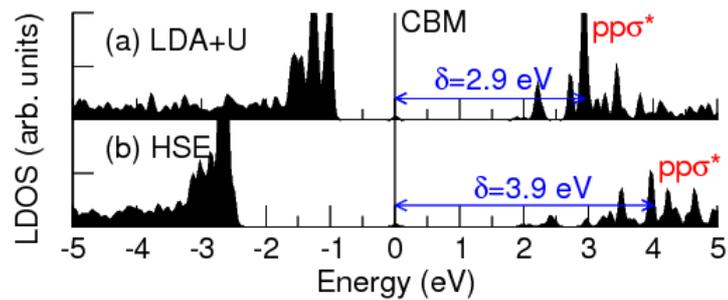

**Figure 6.** (Color online) Calculated LDOS near the O-O bond in the peroxide state of amorphous InGaZnO$_4$ in (a) LDA+U and (b) HSE. The CBM is set to zero in energy level. The Fermi level is located at the CBM.

- 18 -

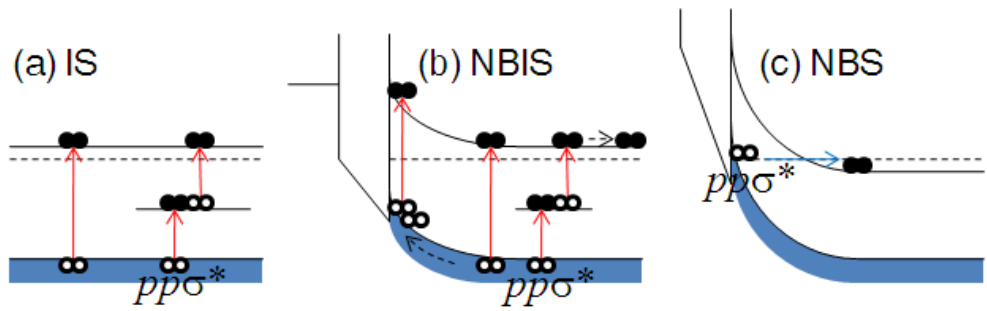

**Figure 7.** (Color online) Schematic illustrations of (a) illumination stress, (b) NBIS, and (c) NBS in band picture. The (red and blue) solid-line arrows indicate the excitation of electron-hole pairs. The dashed-line arrows indicate the drift motion of carriers. The horizontal dashed lines are the Fermi level, and the shadow regions indicate the VBT.